\documentclass[12pt,a4paper,reqno]{article}

\usepackage{amsmath,latexsym,amssymb,amsthm}
\usepackage{graphicx}
\usepackage{amsmath}\usepackage{amssymb}
\usepackage{amsthm}
\usepackage[dvips]{hyperref}

\newcommand{\be}{\begin{equation}}
\newcommand{\ee}{\end{equation}}

\renewcommand{\d}{\text{d}}

\newcommand{\e}{\mathrm{e}}

\newcommand{\m}{\mathbf{m}}
\newcommand{\n}{\mathbf{n}}

\renewcommand{\S}{\mathcal{S}}

\newcommand{\N}{\mathbb{N}}
\renewcommand{\H}{\mathcal{H}}

\renewcommand{\L}{\mathcal{L}}

\newcommand{\<}{\langle}
\renewcommand{\>}{\rangle}

\newcommand{\tr}{\operatorname{tr}}

\theoremstyle{definition}

\theoremstyle{remark}

\numberwithin{equation}{section}

\newcommand{\Pii}{\widetilde{\Pi}}

\begin{document}

\title{\bf{John Bell and the great enterprise}}

\author{Anthony Sudbery$^1$\\[10pt] \small Department of Mathematics,
University of York, \\[-2pt] \small Heslington, York, England YO10 5DD\\
\small $^1$ tony.sudbery@york.ac.uk}

\date{}

\maketitle

\begin{abstract}

 I outline  Bell's vision of the ``great enterprise" of science, and his view that conventional teachings about quantum mechanics constituted a betrayal of this enterprise. I describe a proposal of his to put the theory on a more satisfactory footing, and review the subsequent uses that have been made of one element of this proposal, namely Bell's transition probabilities regarded as fundamental physical processes.

\end{abstract}

\section{Introdution: The Great Enterprise}

John Bell was a scientist. That was a vocation that he followed with great respect, devotion and sense of responsibility. For him, to be a scientist was to participate in the ``great enterprise" \cite{Bell:piddling} of understanding the world we live in; in particular, to be a physicist was to pursue the grand vision of describing the physical world in terms of its ultimate constituents and delineating how those constituents behave. The great enterprise is undertaken according to the scientific method: first, carefully observe and experiment to see how what happens in the world; second, imaginatively construct theories to explain these observations; third, rigorously test these theories by calculating what they predict for the results of further experiments. If these predictions are successful, we can feel, diffidently and tentatively, that we have made progress towards our original goal of truly describing and understanding the world.

When Bell embarked on his career as a physicist, the furthest advances towards the goal of physics were represented by quantum mechanics, as developed and expounded by Bohr, Born and Heisenberg. All the professional training he received followed the teaching of these great men: not only their discoveries, but also their pronouncements on how these discoveries should be regarded, and how future physics should be conducted. Bell was puzzled and dismayed. He felt that everything he was taught constituted a betrayal of the great enterprise: a surrender to the difficulties of the pursuit, and an insistence that there was no alternative but to join the leaders of the field in retreat.

The doctrine which he found all physicists were expected to accept seemed to him to be a distortion of the scientific method. It dismissed, or forgot, the purpose of the method, and held up the method itself as if the very essence of science was contained in the third of these steps: the purpose of physics is to predict the results of experiments. It was a central feature of quantum mechanics, according to the founding fathers, that it could \emph{only} describe the results of experiments. The aim of describing the world apart from experiments was totally and explicitly abandoned. This doctrine was criticised, in a text-book influenced by Bell, as follows:
\begin{quote}
It cannot be true that the sole purpose of a scientific theory is to predict the results of experiments. Why on earth would anyone want to predict the results of experiments? Most of them have no practical use; and even if they had, practical usefulness has nothing to do with scientific enquiry. Predicting the results of experiments is not the \emph{purpose} of a theory, it is a \emph{test} to see if the theory is true. The purpose of a theory is to understand the physical world. (\cite{QMPN}, p. 214)
\end{quote}
In Bell's own words,
\begin{quote}
To restrict quantum mechanics to be exclusively about piddling laboratory operations is to betray the great enterprise.
\end{quote}

Bell's unhappiness with this situation led him to examine the possibility of explaining the results of quantum mechanics in terms of ``hidden variables" --- some way of describing the actual disposition of the material world, regardless of whether any experiments were being done. In the early 1960s it was known that this could be done, following David Bohm's revival in 1952 of a model proposed by Louis de Broglie in 1927. However, although this was known, it was not widely known; indeed, it was generally thought to be impossible because of Pauli's early opposition, to which de Broglie himself surrendered, and a supposed proof by the respected mathematician John von Neumann. But as Bell wrote \cite{Bell:pilot}, ``In 1952 I saw the impossible done". In \cite{Bell:pilot} he analysed the reasons why it continued to be the accepted opinion that hidden variables were impossible, and acknowledged that there were good reasons not to like the de Broglie/Bohm model; Einstein, for example, whom Bell followed in his dissatisfaction with quantum mechanics, found this solution ``cheap". It made sense, therefore, for Bell to look at the full range of possible hidden-variable models; and in doing so he discovered that one particular reason for disliking the de Broglie/Bohm model was unavoidable: any such model would necessarily exhibit \emph{nonlocality}, the feature that distant parts of the model would affect each other instantaneously. This discovery is what Bell is famous for. Most of the papers in this issue of \emph{Quanta} will be devoted to this topic. In this paper, however, I want to focus on one of his later contributions to the project of rescuing the great enterprise. But I should emphasise that there is no substitute for reading Bell's contributions in his own wonderfully elegant and entertaining sentences \cite{Bell:book}.

\section{Beables}

For Bell, the feature of nonlocality, or action at a distance, was no reason to reject a theory. It might be surprising, it might be difficult to reconcile with special relativity, and it might, as it did for Einstein, defy one's presupposition of what a scientific description of the world should look like; but this is outweighed by the virtue of actually giving a description of the world, independent of human beings and ``piddling laboratory operations" \cite{Bell:piddling}. In conversation, Bell would emphasise that he would encourage anyone working on a theory with this overriding virtue. He put aside his own opinions as to whether the work was likely to be successful; the important thing was to get physicists thinking in a healthy, ``professional" \cite{Bell:beables} way. And he was enormously helpful and supportive: I remember, at a conference in 1987, diffidently giving him the manuscript of a paper at the end of one afternoon. Despite attending an alcoholic reception that evening, he sought me out the next morning to give me detailed comments.

Bell's hostility to the official version of quantum mechanics, as preached in nearly all university physics courses, is emblazoned in the two words of the title of his paper ``Against `measurement'" \cite{Bell:piddling}. Another key word to which he took exception is ``observable". This is the only word available in the official theory to refer to properties of physical objects; it insists that the only quantities that can have any place in a physical theory are those which can be measured, or \emph{observed}. In line with his conviction that a theory should describe physical objects themselves, regardless of their relation to human observers, Bell proposed \cite{Bell:subjobj} to replace the word ``observable" by a new coinage of ``beable"
 to emphasise the autonomous existence of the quantities in question. 

In the Bohm/de Broglie theory of non-relativistic many-particle quantum mechanics, the beables are the positions of the particles. This might be one of the reasons why the theory is not universally liked. A fundamental feature of the conventional theory is that all the quantities of classical mechanics, i.e. functions on phase space, have quantum counterparts which enter the theory on an equal footing, as they do in Hamiltonian mechanics. This symmetry under canonical transformations, becoming unitary symmetry in quantum mechanics, is widely regarded as very attractive. It is explicitly broken in de Broglie/Bohm theory. Bell defended this, arguing that all actual observations in experimental physics come down to measurements of position (of pointers in instruments, for example). This strikes me as dubious, and anyway it represents a reliance on experimental considerations which is curiously inappropriate in the author of \emph{Against ``measurement"}. However, in the paper in which I now want to focus \cite{Bell:beables}, Bell addressed a more serious problem: de Broglie/Bohm theory is non-relativistic, and is a theory of a finite number of particles. On the other hand, the fundamental theories which Bell was seeking as an elementary particle physicist would have to be relativistic, and they would have to be theories of fields, not particles. Hence his title: \emph{Beables for quantum field theory}.

In this paper Bell tackled only one of these desiderata. He proposed a quantum theory of fields with a clearly defined set of beables, quantities with privileged ontological status, and with dynamics which reproduced the predictions of orthodox quantum theory, just as the Bohm/de Broglie equation of motion reproduces the predictions of the Schr\"odinger equation. But his theory is not relativistic. He assumes an absolute distinction between space and time, in which time is continuous but space is taken to be a discrete three-dimensional lattice $\mathcal{L}$. The beables of the theory are the total numbers of fermions at the points of the lattice. These fermion numbers are the eigenvalues of the field operators 
\[
B(x) = \sum_i \psi_i(x)^\dagger\psi_i(x), \qquad x\in\L
\]
where $\psi_i$ is a particular variety of Dirac field, labelled by $i$, and the sum is over all such varieties. Thus the actual situation of the world, is given by the set of integers $\{F(x):\; x\in\L\}$ representing numbers of fermions at all points of the lattice. Bell does not consider it necessary or even possible for the numbers of different types of fermion to be beables, since interacting fields will not all commute. He regards microscopic details such as these as ``entirely redundant. What is essential is to be able to define the positions of things", by which he seems to mean things that we would recognise in our macroscopic world. His only elucidation of these ``things" is that they should include ``positions of instrument pointers or (the modern equivalent) of ink on computer output". This is not intended to be exhaustive: it is not that these are the only beables but that the beables must \emph{at least} include the positions of instrument pointers.

\section{Transition probabilities}

In Bell's model the state of the world at each time $t$ is given by an element $\n(t)$ of the discrete set of functions $\n:\L\to\N$ ($\N$ being the set of non-negative integers), so each $\n$ is a set of non-negative integers, one for each lattice point. The change of this state with time is governed, as in de Broglie/Bohm theory, by a time-dependent element $|\Psi(t)\>$ of a Hilbert space spanned by eigenvectors of the field operators. We ae used to calling $|\Psi(t)\>$ a ``state vector", but in this theory that terminology is misleading: $|\Psi(t)\>$ does not describe a state of the world, but something that governs change in the state of the world. Let us call it the ``pilot vector", in memory of the pilot wave of dBB theory. However, the value of $|\Psi(t)\>$ is a fact about the world, and Bell therefore considered $|\Psi(t)\>$, as well as $\n(t)$, to be a beable. The complete specification of the world at time $t$ is then given by the pair $(\n(t),|\Psi(t)\>)$.
 
The way that the world changes in time is an adaptation of the evolution equations of de Broglie/Bohm theory. As in that theory, the pilot vector $|\Psi(t)\>$ evolves according to the Schr\"odinger equation with a Hamiltonian $H$:
\[
i\hbar\frac{\d}{\d t}|\Psi(t)\> = H|\Psi(t)\>.
\]
Because the possible values of $\n(t)$ are a discrete set, however, the change from one value to another is stochastic: Bell postulates that if, at time $t$, the value of the total fermion number distribution is $\m$, then the probability that at time $t + \delta t$ its value has changed to $\n$ is $w_{\m\n}\delta t$ where the transition probability $w_{\m\n}$ is given by
\be\label{Bell}
w_{\m\n} = \begin{cases} \frac{2\text{Re}[(i\hbar)^{-1}\<\n|H|\m\>\overline{c_\n}c_\m]}{\<\psi(t)|P_\m|\psi(t)\>} &\text{if this is } \ge 0\\
                       0                                                  &\text{if it is negative}
         \end{cases}
\ee 
where $P_\m$ is the projection onto the subspace of simultaneous eigenstates of the local occupation number operators $B(x)$ in which the eigenvalue of $B(x)$ is $\m(x)$. Bell then shows that this joint time development is consistent with the probabilities given by the Born rule in the same sense as in dBB theory: if the Born rule holds at some initial time, i.e. the value of the total fermion number distribution at that time is given probabilistically so that the probability of the distribution $\m$ is $\<\Psi(0)|P_\m|\Psi(0)\>$, then this remains true at all subsequent times.

Bell found the stochastic nature of this time development ``unwelcome"; he suspected that it was purely a consequence of his artificial assumption of a discrete lattice of points of space, and that it would ``go away in some sense in the continuum limit". Indeed, it was shown by Vink \cite{Vink} and myself \cite{determlimit}, working independently, that a stochastic model of a particle on a one-dimensional lattice, modelled on this theory of Bell's, did become the deterministic dBB theory in the continuum limit. Bell's unease arose from his respect for the time-reversal invariance of both quantum and classical mechanics, in the forms of Schr\"odinger's equation and Newton's equations of motion. Others, however, have welcomed both stochastic elements in fundamental theory, as reflecting our actual experience of quantum phenomena, and non-invariance under time reversal, as reflecting the true nature of time. (``Others" here possibly means just myself \cite{verdammte}.)

The transition probabilities introduced by Bell can be used in a wide variety of theories, not restricted to those which postulate a special class of quantities which are ``beable". In general, consider a theory which supposes that there is a true description of the world by means of a vector $|\Psi(t)\>$ which evolves according to a Schr\"odinger equation with Hamiltonian $H$, and that there is some reason to give special consideration to one of the components of this vector in a decomposition given by special subspaces $\S_n$ of the Hilbert space $\H$, known as \emph{viable} subspaces. Thus $\H$ is the orthogonal direct sum of the subspaces $\S_n$, and any $|\Psi\> \in \H$ can be written
\[
|\Psi(t)\> = \sum_n|\psi_n(t)\> \qquad \text{with} \qquad |\psi_n(t)\>\in\S_n.
\]
Then $|\psi_n(t)\> = P_n|\Psi(t)\>$ where $P_n$ is the orthogonal projection onto the subspace $\S_n$. As $|\Psi(t)\>$ changes in accordance with the Schr\"odinger equation, the components $|\psi_n(t)\> = P_n|\Psi(t)\>$ will also change inside their respective viable subspaces; but in addition to this, the spotlight shining on the component with special status will also move stochastically from one subspace to another. This stochastic change is given by Bell's transition probabilities: if the special component is in subspace $\S_m$ at time t, then the probability that it has moved to subspace $\S_n$ by time $t + \delta t$ is $w_{mn}\delta t$ where $w_{mn}$ is given by \eqref{Bell}. If the viable subspaces $\S_n$ are themselves changing with time, then there is an extra term in this equation (\cite{BacciaDickson, verdammte}).

I will refer to theories with this structure as ``generalised Bell-type theories". All such theories share the property proved by Bell for his version of quantum field theory: the transition probabilities \eqref{Bell} guarantee the continuing validity of the Born rule if it is valid initially. They are not uniquely determined by this requirement: there is a range of possible transition probabilities with the same property \cite{BacciaDickson}. However, Bell's formula is uniquely natural in applications to decay \cite{verdammte} and measurement processes \cite{singleworld, Hollowood:classical}: it ensures that the underlying direction of change in such processes is always forwards, without intermittent reversals (decay products, for example, recombining to reconstitute the unstable decaying state).

In many such theories the special status of the highlighted component $|\psi_n(t)\>$ is ontological; only this component describes the actual state of the world, and the function of the overall vector $|\Psi(t)\>$ is to act as a pilot, guiding the discontinuous quantum transitions of the world. Such theories are liable to face a \emph{preferred basis} problem: what defines the viable subspaces $\S_n$? Bell formulated the concept of \emph{beables} precisely to give an answer to this problem: the viable subspaces are the eigenspaces of beables. In the original dBB theory, the beables are the particle positions. We have already noted on the one hand the unease this arouses because of its violation of symmetry under canonical transformations, and on the other hand Bell's defence of it on the grounds that ultimately all observations are of position. 

In Bell's theory in \cite{Bell:beables}, the beables are the total fermion numbers at each point in the lattice of space. At first sight it seems reasonable that these should have fundamental status, but this is thrown into doubt by the Unruh effect, according to which the number of particles present in a region of space depends on a frame of reference as soon as one moves away from frames in constant relative motion.

This preferred-basis problem is also often thought to arise in the ``many-worlds" interpretation of quantum mechanics. That theory has only the universal state vector $|\Psi(t)\>$ and does not single out a component of that vector as describing the actual world. Nevertheless, if the vector $|\Psi(t)\>$ is regarded as describing many worlds, all of which are real, then some commentators, including Bell \cite{Bell:sixworlds}, demand that there should be a specification of which vectors or subspaces can describe ``worlds". However, this is not a problem in Everett's original version \cite{Everett} of this interpretation but only arises when too much weight is placed on the expository terminology of ``worlds" (\cite{QMPN} p. 221). Even with this terminology, the components of the state vector $|\Psi(t)\>$ which describe worlds can be determined by the structure of $|\Psi(t)\>$ itself \cite{Wallace:multiverse} and need no independent definition.

Other generalised Bell-type theories have no preferred-basis problem, defining the viable subspaces purely in terms of the pilot vector $|\Psi(t)\>$. In the (now largely discarded) modal interpretation, in which the universe is divided into two systems so that the Hilbert space $\H$ is a tensor product of two factors, the viable components are defined by the Schmidt decomposition of $|\Psi(t)\>$ with respect to this structure. 

The format of a generalised Bell-type theory is appropriate to describe the changing experience of a sentient subsystem of the physical universe \cite{logicfuture}. This can be done in the context of Everettian theory, in which the universe is described by a single time-dependent state vector $|\Psi(t)\>$, and nothing else. We know that the universe has sentient subsystems, each of which is capable of experiences relating to the rest of the universe. I am myself such a subsystem. I have various possible experiences, for each of which there is a set of physical states of my body in which I have the experience. Since I can distinguish between the experiences (if I couldn't they wouldn't be different experiences), it seems to be in keeping with quantum mechanics that the corresponding states form a set of orthogonal subspaces $\S^{\text{me}}_n$ of my Hilbert space $\H_{\text{me}}$. These subspaces then define a set of subspaces $\S_n = \S_n^{\text{me}}\otimes\H_{\text{rest}}$ of the universal Hilbert space $\H = \H_{\text{me}}\otimes\H_{\text{rest}}$. The changes in my experience then constitute transitions between these subspaces. Bell's formula \eqref{Bell} gives the probabilities of these transitions, subject to a universal state vector and a universal Hamiltonian.

This formalism can also be used \cite{singleworld, Hollowood:classical} to model the progress of a quantum measurement. 

\section{Histories}

Once a significant set of preferred subspaces has been identified for each time $t$, a generalised Bell-type theory makes it possible to calculate probabilities for \emph{histories} of the system. It is usual, and convenient, to consider only a discrete set of times $t_1,\ldots,t_f$; then a \emph{history} of the system is a sequence $(\S_1,\ldots,\S_f)$ where each $\S_i$ is a closed subspace of the Hilbert space of the system, or equivalently a sequence of projection operators $h = (\Pi_1,\ldots,\Pi_f)$ where $\Pi_i$ is the projection onto $\S_i$. Such histories are the fundamental concepts in the \emph{consistent histories} interpretation of quantum mechanics \cite{Griffiths:book}, in which the probability of the history $h$ is taken to be
\be\label{consprob}
P(h) = \tr[\Pii_1\ldots\Pii_{f-1}\Pii_f\Pii_{f-1}\ldots\Pii_1]
\ee
where 
\[
\Pii_i = \e^{iHt/\hbar}\Pi_i\e^{-iHt/\hbar}.
\]
This probability can be obtained from the Copenhagen interpretation of quantum mechanics by assuming that at each time $t_i$ there is a measurement of an observable whose eigenspaces include $\S_i$, and applying the projection postulate after each measurement. Then $P(h)$ is the probability that this sequence of measurements has the results corresponding to the subspaces $\S_1,\ldots\S_n$. It can be written
\[
p(h) = \tr[C_h C_h^\dagger]\notag
\]
where $C_h$ is the \emph{history operator}
\be\label{historyop}
C_h = \Pii_1\cdots\Pii_f.
\ee

In general, these probabilities will not be consistent with the following natural requirement. Suppose two histories $h_1$ and $h_2$ differ only in the subspaces $\S^{(1)}_i$ and $\S^{(2)}_i$ which they assign to time $t_i$, and that these subspaces are orthogonal. We can consider a third history $h_1\lor h_2$ in which the subspace at time $t_i$ is the direct sum $\S_1\oplus\S_2$. In terms of measurement, this describes a result of the measurement at time $t_i$ which was either the result corresponding to $\S_1$ or that corresponding to $\S_2$; so $h_1\lor h_2$ relates that the history of the system was either $h_1$ or $h_2$. We expect that the corresponding probabilities should satisfy
\[
P(h_1\lor h_2) = P(h_1) + P(h_2).
\]
In particular, if $\S_1\oplus\S_2 = \H$, we expect that $\S_1$ and $\S_2$ are an exhaustive set of possibilities at time $t_i$ and so
\[
P(h_1) + P(h_2) = P(h')
\]
where $h'$ is the history which coincides with $h_1$ and $h_2$ at all times except $t_i$, but does not say anything about time $t_i$.
However, these equations will not in general be true. A condition which guarantees them is
\be\label{CH}
\tr[C_{h_1} C_{h_2}^\dagger] = 0.
\ee
A set of histories is said to be \emph{consistent} (or \emph{decoherent}) if this condition holds true for every pair of different histories in the set.

This is not an issue in generalised Bell-type theories. It would be an issue if the transition probabilities \eqref{Bell} were supposed to apply for transitions to any subspace in the Boolean algebra generated by the subspaces $\S_i$, but that would not be in accord with the basic presuppositions of such a theory. To take a linear sum of the preferred subspaces as having the same status as those subspaces would be to assume that the system could exist in a superposition of states from the preferred subspaces, whereas the philosophy of these theories is that such superpositions are not actual states. In the theory of sentient experience, for example, a sum of experience states describing different experiences is not an experience state (a sum of eigenvectors with different eigenvalues is not an eigenvector). Thus the state of the system can be in a subspace $\S_1\oplus\S_2$ only if it is in the subset $\S_1\cup\S_2$, and the appropriate probability is 
\[
P(\S_1\oplus\S_2) = P(\S_1) + P(\S_2).
\]

In \cite{logicfuture}, where probabilities are identified with truth values, and a history is formed by logical operations from single-time propositions, the probability of a history was taken to be given by the usual formula \eqref{consprob}. For the development of a satisfactory logic, it was then found to be necessary to make the consistent-histories assumption \eqref{CH} (in a somewhat weaker version). I now think that this was a mistake. If Bell's transition probabilities had been used to define the truth value (= probability) of a history rather than \eqref{consprob}, there would have been no need for a subsidiary assumption, and the logic could have been developed in much greater generality.

\section{Conclusion}

John Bell never lost sight of the great enterprise of science. He rejected the narrow scepticism and pessimism in the reaction of the founding fathers of quantum mechanics to the difficulties which they encountered, and the instrumentalist view of physics which became the dogma in which all physics students were indoctrinated. His own most famous and influential work only served to emphasise the difficulties in the way of understanding quantum mechanics as he thought physical theories should be understood. Nevertheless, he persevered in the search for such an understanding. His concept of ``beables" has become a standard tool for those seeking to understand and develop quantum mechanics, and deserves deeper philosophical analysis. The transition probabilities that he formulated as a component of theories of such beables are a lasting legacy of his search, and have proved valuable even to those who do not share his vision of what a satisfactory physical theory should be like.


\end{document}